\documentclass{elsart}
\usepackage{amsmath,amssymb,amscd}
\journal{Physics Letters B}
\allowdisplaybreaks
\newcommand{\al}{\alpha}
\newcommand{\be}{\beta}

\newcommand{\ep}{\epsilon}

\newcommand{\la}{\lambda}

\newcommand{\vp}{\varphi}
\newcommand{\ze}{\zeta}
\newcommand{\La}{\Lambda}

\newcommand{\bx}{\mathbf{x}}

\newcommand{\bs}{\mathbf{s}}

\newcommand{\bz}{\mathbf{z}}

\newcommand{\tH}{\widetilde{H}}
\newcommand{\tcH}{\widetilde{\cH}}

\newcommand{\ssH}{\mathsf{H}}

\newcommand{\cH}{{\mathcal H}}

\newcommand{\cP}{{\mathcal P}}
\newcommand{\cQ}{{\mathcal Q}}

\newcommand{\cS}{{\mathcal S}}
\newcommand{\cT}{{\mathcal T}}

\newcommand{\pa}{\partial}

\def\ket#1{|#1\rangle}

\newcommand{\ms}{\mspace{1mu}}
\newcommand{\iu}{{\mathrm i}}
\begin{document}
\begin{frontmatter}
\title{Solvable scalar and spin models with near-neighbors interactions}
\author{A. Enciso},
\author{F. Finkel},
\author{A. Gonz{\'a}lez-L\'opez\corauthref{cor}},
\corauth[cor]{Corresponding author. E-mail: \texttt{artemio@fis.ucm.es}.}
\author{M.A. Rodr{\'\i}guez}
\address{Depto.~de F{\'\i}sica Te\'orica II, 
Universidad Complutense, 
28040 Madrid, Spain}
%
%
\date{September 2, 2004}
\begin{abstract}
We construct new solvable rational and trigonometric spin models
with near-neigh\-bors interactions by an extension of the Dunkl
operator formalism. In the trigonometric case we obtain a finite
number of energy levels in the center of mass frame, while the
rational models are shown to possess an equally spaced infinite
algebraic spectrum. For the trigonometric and one of the rational
models, the corresponding eigenfunctions are explicitly computed.
We also study the scalar reductions of the models, some of which
had already appeared in the literature, and compute their
algebraic eigenfunctions in closed form. In the rational cases,
for which only partial results were available, we give concise
expressions of the eigenfunctions in terms of generalized Laguerre
and Jacobi polynomials.
\end{abstract}
\end{frontmatter}
%
%
Calogero--Sutherland (CS) models are one of the most extensively
studied types of exactly solvable and integrable quantum
Hamiltonians describing a system of $N$ particles in one dimension
with long-range two-body
interactions~\cite{Ca71,Su71,Su72,OP83,DV00}. Apart from their
intrinsic mathematical relevance~\cite{Ka95,LV96,BF97}, these
models arise naturally in many different fields, such as
Yang--Mills theories~\cite{GN94,HP98}, quantum Hall
liquids~\cite{AI94}, random matrix theory
\cite{Su71b,Su71c,SLA94,TSA95}, propagation of solitons
\cite{Po95}, fractional statistics and
anyons~\cite{Po89,Ha94,Ha95,CL99,Sh01}, etc. Several
generalizations of CS models to particles with internal degrees of
freedom (``spin'') have been developed over the last decade using
two main approaches, namely the supersymmetric
formalism~\cite{BTW98,DLM01,Gh01} and the Dunkl operator
method~\cite{Du89,Po92,Ya95,Ba96,FGGRZ01,FGGRZ01b}. Spin CS models
are intimately connected with integrable spin chains with
long-range position-dependent interactions, like the celebrated
Haldane--Shastry spin chain~\cite{Ha88,Sh88}. Indeed, when the
coupling constant of a spin CS model tends to infinity the
particles ``freeze'' in the classical equilibrium positions of the
scalar part of the potential, thus giving rise to a spin chain of
Haldane--Shastry (HS) type. This mechanism, usually called the
``freezing trick'', was first used by Polychronakos~\cite{Po93} to
construct the first integrals of the original HS spin chain by
taking the large coupling constant limit of the corresponding
integrals of the Sutherland spin model. These ideas have also been
successfully applied to construct integrable and exactly solvable
spin chains of HS type from other spin CS
models~\cite{Po94,BPS95,YT96,EFGR04}.

In a recent paper~\cite{JK99}, Jain and Khare proposed new
solvable versions of the original Calogero and Sutherland scalar
models featuring near-neighbors interactions. These new models are
closely related to the so-called short-range Dyson
models~\cite{GJ98,BGS99} in random matrix theory, in that the
square of the ground-state wavefunction of the many-body system
coincides with the joint probability distribution function for
eigenvalues of the corresponding short-range Dyson model. Several
generalizations of the Jain and Khare models appeared in a
subsequent paper~\cite{AJK01}, including their extension to the
$BC_N$ root system and to higher dimensions. The previous
publications open a number of interesting questions, such as the
existence of other solvable scalar models with near-neighbors
interactions, or the construction of spin models with near-neighbors interactions
and their corresponding spin chains of
Haldane--Shastry type. These short-range spin chains with position-dependent
interactions are of particular significance, since they
occupy an intermediate position between the well-known Heisenberg
chain (short-range, position-independent interactions) and the
usual HS-type spin chains (long-range, position-dependent
interactions). A first step in this direction is the recent work
by Deguchi and Ghosh~\cite{DG01}, in which several spin $1/2$
models related to the scalar models of Jain and Khare were
introduced and partially solved using the supersymmetric
formalism. These authors also pointed out how to obtain the spin
chains corresponding to these models by applying Polychronakos's
freezing trick.

In this letter we present three new families of solvable scalar
and spin $N$-body models with near-neighbors interactions. By contrast
with Ref.~\cite{DG01}, our approach is based on a modification
of the Dunkl operator formalism and provides a wide range
of totally explicit solutions for all values of the spin.
The potentials of these spin models are given by
\begin{subequations}\label{Vs}
\begin{align}
V_1&=2a^2\sum_i\cot(x_i-x_{i-1})\cot(x_i-x_{i+1})\notag\\
&\hphantom{{}=\omega^2 r^2+\sum_i\frac{b(b-1)}{x_i^2}}
+2a\sum_i\sin^{-2}(x_i-x_{i+1})\,(a-S_{i,i+1})\,,\label{V00}\\
V_2&=\omega^2 r^2+\sum_i\frac{b(b-1)}{x_i^2}
+\sum_i\frac{8a^2x_i^2}{(x_i^2-x_{i-1}^2)(x_i^2-x_{i+1}^2)}\notag\\
&\hphantom{{}=\omega^2 r^2+\sum_i\frac{b(b-1)}{x_i^2}}
+4a\sum_i\frac{x_i^2+x_{i+1}^2}{{(x_i^2-x_{i+1}^2)}^2}
\,(a-S_{i,i+1})\,,\label{V0-}\\
V_3&=\omega^2 r^2+\sum_i\frac{2a^2}{(x_i-x_{i-1})(x_i-x_{i+1})}+
\sum_i\frac{2a}{(x_i-x_{i+1})^2}\,(a-S_{i,i+1})\,,\label{V--}
\end{align}
\end{subequations}
where $r^2=\sum_i x_i^2$ and $a,b>1/2$. Here and in what follows
the sums run from $1$ to $N$, and we are identifying $x_{N+1}$
with $x_1$. The operators $S_{ij}$ permute the spin coordinates of
the particles $i$ and $j$. In other words, if
$\ket{s_1,\dots,s_N}$ (with $-M\leq s_i\leq M$, $M$ being a
half-integer) is an element of the basis of the spin space~$\cS$,
then
\[
S_{ij}\ket{\dots,s_i,\dots,s_j,\dots}=\ket{\dots,s_j,\dots,s_i,\dots}.
\]
Note that $S_{ij}$ can be expressed in terms of the fundamental
$\mathrm{SU}(2M+1)$ generators $S_i^a$, $a=1,\dots,4M(M+1)$, as
$S_{ij}=1/(2M+1)+\sum_a S_i^aS_j^a$.

The spin potentials~\eqref{Vs} reduce to solvable scalar
potentials by setting $S_{i,i+1}$ to $1$. In particular, the
scalar reductions of the potentials~\eqref{V00} and~\eqref{V--}
are the models introduced by Jain and Khare. The
potentials~\eqref{V00} and~\eqref{V--} with $M=1/2$ are similar to
the spin $1/2$ potentials introduced in Ref.~\cite{DG01}, but
differ from them by a spin-dependent term. The spin
model~\eqref{V0-}, as well as its scalar reduction, are both
completely new. There is also a hyperbolic version of the
trigonometric potential~\eqref{V00}, obtained by replacing $x_k$
by $\iu\ms x_k$ and $V_1$ by $-V_1$.

The starting point in the solution of the models~\eqref{Vs} is the
introduction of the following second-order differential-difference
operators, which play the same role as the quadratic combinations
of Dunkl operators in the construction of CS models with
spin~\cite{FGGRZ01,FGGRZ01b}:
\begin{subequations}\label{Ts}
\begin{align}
T_1&=\sum_iz_i^2\pa_i^2+2a\sum_i\frac1{z_i-z_{i+1}}\,(z_i^2\pa_i-z_{i+1}^2\pa_{i+1})\hspace{6em}\notag\\
&\hspace{10em}-2a\sum_i\frac{z_iz_{i+1}}{(z_i-z_{i+1})^2}\,(1-K_{i,i+1})\,\label{T1}\\
T_2&=\sum_iz_i\pa_i^2+2a\sum_i\frac1{z_i-z_{i+1}}\,(z_i\pa_i-z_{i+1}\pa_{i+1})\notag\\
&\hspace{10em}-a\sum_i\frac{z_i+z_{i+1}}{(z_i-z_{i+1})^2}\,(1-K_{i,i+1})\,,\label{T2}\\
T_3&=\sum_i\pa_i^2+2a\sum_i\frac1{z_i-z_{i+1}}\,(\pa_i-\pa_{i+1})\notag\\
&\hspace{10em}-2a\sum_i\frac1{(z_i-z_{i+1})^2}\,(1-K_{i,i+1})\,.\label{T3}
\end{align}
\end{subequations}
Here $\pa_i\equiv\pa_{z_i}$, the operator $K_{ij}$ permutes the
variables $z_i$ and $z_j$, and we are again identifying $z_{N+1}$
with $z_1$. We shall also need in what follows the first-order
operators
\begin{equation}\label{J0J-}
J_-=\sum_i \pa_i,\qquad J_0=\sum_i z_i\pa_i\,.
\end{equation}
If $\Phi=\sum_\bs f_\bs(\bz)\ket\bs$, where $\bz=(z_1,\dots,z_N)$
and $\bs=(s_1,\dots,s_N)$, is a state totally symmetric under
permutations of both the spatial and spin coordinates of each
particle, then $K_{ij}\Phi=S_{ij}\Phi$. Hence
$T_\ep\Phi=T_\ep^*\Phi$, $\ep=1,2,3$, where
$T^*_\ep=T_\ep|_{K_{i,i+1}\to S_{i,i+1}}$. The Hamiltonian
$H_\ep=-\sum_i\pa_{x_i}^2+V_\ep$ of each of the models~\eqref{Vs}
can be obtained by applying a suitable gauge transformation and
change of variables to a linear combination
\begin{equation}\label{tHep}
\tH_\ep=c\,T^*_\ep+c_- J_-+c_0 J_0+E_0\,.
\end{equation}
More precisely, we can write
\begin{equation}\label{Hep}
H_\ep=\mu\cdot\tH_\ep\big|_{z_i=\ze(x_i)}\cdot\mu^{-1}\,,
\end{equation}
where the constants $c$, $c_-$, $c_0$, $E_0$, the gauge factor
$\mu$, and the function $\ze$ are given in each case by
\begin{subequations}\label{coefs}
\begin{align}
&1)&& c=4,\quad c_-=0,\quad c_0=4(1-2a),\quad E_0=2Na^2,\notag\\
&&& \mu=\prod_i\sin^a(x_i-x_{i+1}),\quad \ze(x)=\e^{\pm2\iu x},\label{coefs1}\\
&2)&& c=-4,\quad c_-=-2(2b+1),\quad c_0=4\omega,\quad E_0=N\omega(4a+2b+1),\qquad\notag\\
&&& \mu=\e^{-\frac\omega2\,r^2}\prod_i(x_i^2-x_{i+1}^2)\vphantom{x}^a\,x_i^b,
\quad \ze(x)=x^2,\label{coefs2}\\
&3)&& c=-1,\quad c_-=0,\quad c_0=2\omega,\quad E_0=N\omega(2a+1),\notag\\
&&& \mu=\e^{-\frac\omega2\,r^2}\prod_i(x_i-x_{i+1})^a,\quad \ze(x)=x.\label{coefs3}
\end{align}
\end{subequations}

The key idea in our approach to the solution of the
models~\eqref{Vs} is to find an increasing sequence of
finite-dimensional linear spaces
$\tcH^0_\ep\subset\tcH^1_\ep\subset\cdots$ invariant under the
gauge Hamiltonian~$\tH_\ep$. Indeed, by Eq.~\eqref{Hep} this
implies that the corresponding Hamiltonian $H_\ep$ can be
diagonalized in each of its invariant subspaces
$\cH^n_\ep=\mu\ms\tcH^n_\ep\big|_{z_i=\ze(x_i)}$, $n=0,1,\dots$.
The operators~\eqref{Ts} and~\eqref{J0J-} preserve the space
$\cP^n$ of polynomials in $\bz$ of total degree at most $n$, for
all non-negative integer values of $n$. In our recent work on spin
CS models~\cite{FGGRZ01,FGGRZ01b}, the operators $T_{\textsc{cs}}$ analogous to
$T_\ep$ also leave $\cP^n$ invariant, and in addition commute with the
total symmetrizer under particle permutations~$\La$. This
guarantees that the corresponding operators $T_{\textsc{cs}}^*$ preserve the
space of completely symmetric spin functions
$\La(\cP^n\otimes\cS)$. In the present case, however, the
operators~\eqref{Ts} do not commute with $\La$, and hence it is not clear
\emph{a priori} whether the operators $T_\ep^*$ leave invariant any
finite-dimensional space of spin functions. In fact, since $J_-$
and $J_0$ obviously commute with $\La$, the results of Jain and
Khare~\cite{JK99} for the scalar case indicate that the operators
$T^*_1$ and $T^*_3$ possess at least ``trivial'' invariant spaces
of the form $\cQ\otimes(\La\cS)$, where $\cQ$ is a
finite-dimensional subspace of the space of totally symmetric
polynomials in $\bz$. As it turns out, each of the operators
$T_\ep^*$ preserves a nontrivial subspace
$\cT^n_\ep\subset\La(\cP^n\otimes\cS)$ for all $n$, namely
\begin{subequations}\label{cTs}
\begin{align}
\cT^n_1 &=\big\langle f(\tau_1,\tau_{N-1},\tau_N)\La\ket s,g(\tau_{N-1},\tau_N)\La(z_1\ket s),
q(\tau_1,\tau_N)\La(z_1\cdots z_{N-1}\ket s)\quad\notag\\
&\hphantom{{}=\big\langle f(\tau_1,\tau_2,\tau_3)\La\ket s,} {}\mid f_{11}=f_{N-1,N-1}=g_{N-1,N-1}=q_{11}=0\big\rangle\,,\label{cT1}\\
\cT^n_2 &=\big\langle f(\tau_1,\tau_2,\tau_N)\La\ket
s,g(\tau_1,\tau_N)\La(z_1\ket s)
\mid{}f_{22}=f_{NN}=g_{NN}=0\big\rangle\,,\label{cT2}\\
\cT^n_3 &=\big\langle f(\tau_1,\tau_2,\tau_3)\La\ket s,g(\tau_1,\tau_2,\tau_3)\La(z_1\ket s),
h(\tau_1,\tau_2)\La(z_1^2\ket s),\notag\\
&\hphantom{{}=\big\langle f(\tau_1,\tau_2,\tau_3)\La\ket s,} h(\tau_1,\tau_2)\La(z_1z_2\ket{s'})\mid f_{33}=g_{33}=0\big\rangle\,.
\label{cT3}
\end{align}
\end{subequations}
The proof of this statement, which is crucial for what follows,
will appear in a separate publication. In Eq.~\eqref{cTs},
\[
\tau_k=\sum\limits_{i_1<\cdots<i_k}z_{i_1}\cdots z_{i_k}
\]
is the $k$-th elementary symmetric polynomial, $f$, $g$, $h$, and $q$ are
polynomials of total degree in $\bz$ less than or equal to $n$,
$n-1$, $n-2$, and $n-N+1$, respectively, and (for instance)
$f_k=\pa f/\pa\tau_k$. The spin states $\ket s\in\cS$ are
arbitrary, while $\ket{s'}$ denotes a state such that the sum
$\sum_i\ket{s'_{i,i+1}}$ is totally symmetric, where
$\ket{s'_{ij}}$ is defined by
$\La\big(z_1z_2\ket{s'}\big)=\sum_{i<j}z_iz_j\ket{s'_{ij}}$. Note
that the spaces~\eqref{cTs} are also preserved by the operator~$J_0$.
Similarly, the operator $J_-$ leaves the space $\cT^n_3$
invariant, while it preserves $\cT^n_1$ and $\cT^n_2$ provided
that $f_{N-1}=f_N=g_{N-1}=g_N=q=0$ and $f_N=g_N=0$, respectively.
From Eqs.~\eqref{tHep} and~\eqref{coefs}, it easily follows that the
gauge Hamiltonians $\tH_\ep$ preserve the spaces $\tcH^n_\ep$
given by
\begin{equation}\label{tcHs}
\tcH^n_1=\cT^n_1\,,\quad
\tcH^n_2=\cT^n_2\big|_{f_N=g_N=0}\,,\quad
\tcH^n_3=\cT^n_3\,.
\end{equation}
By diagonalizing the Hamiltonian $H_\ep$ in its corresponding
invariant spaces $\cH^n_\ep$, $n=0,1,\dots$, one can in principle
construct an infinite sequence of exact eigenvalues and
eigenfunctions, which shall be referred to as ``algebraic'' in
what follows. We have found in this way all the algebraic
eigenvalues of the spin models~\eqref{Vs}. We also present
explicit expressions for the corresponding algebraic
eigenfunctions, with the only exception of the spin eigenfunctions of the
model~\eqref{V--} not factorizing as $\mu
f(\tau_1,\tau_2,\tau_3)\La\ket s$. In particular, we obtain
all the algebraic eigenfunctions of the scalar reductions of
the spin potentials~\eqref{Vs}, thus considerably extending the
results of Refs.~\cite{JK99} and~\cite{AJK01}. It should be noted, however, that
the point spectrum could possibly include additional eigenvalues
and eigenfunctions which are not algebraic. We shall now discuss
in more detail each of the models~\eqref{Vs}.

\emph{Case 1.} The Hamiltonian $H_1$ commutes with the total
linear momentum $P=-\iu\sum_k\pa_{x_k}$, so that it admits a basis
of eigenfunctions with well-defined total momentum. In fact, since
in this case $\tau_N^l=\exp\big(\pm 2\iu\ms l\sum_k\!x_k\big)$,
multiplying an eigenfunction of $H_1$ with energy $E$ and total
momentum $p$ by $\tau_N^l$ simply ``boosts'' its energy and total
momentum. We shall take advantage of this fact to ``normalize'' the
algebraic eigenfunctions to zero total momentum. It easily follows
from Eq.~\eqref{cT1} that when this normalization is performed one
obtains only a finite number of eigenfunctions of $H_1$. In the
scalar case, there are exactly four eigenfunctions with zero momentum, namely
\begin{equation}\label{psis1}
\begin{aligned}
\psi_0&=\mu\,,\qquad \psi_{1,2}=\mu\sum_i\bigg\{\begin{matrix}\cos\\\sin\end{matrix}\bigg\}
\big(2(x_i-X)\big)\,,\\
\psi_3&=\mu\bigg[\frac{Na}{2a+1}+\sum_{i<j}\cos(x_i-x_j)\bigg]\,,
\end{aligned}
\end{equation}
where $X=\frac1N\sum_ix_i$ is the center of mass coordinate. Their
respective energies are $E_0$ (ground state),
$E_{1,2}=E_0+4(2a-1+1/N)$, and $E_3=E_0+8(2a+1)$. These are
essentially the solutions found in~\cite{EGKP02}.

In the spin case, to each scalar eigenfunction~\eqref{psis1} there
correspond $\binom{2M+N}N$ factorized solutions of the form
$\Psi^{(0)}_n=\psi_n\La\ket s$, where $\ket s\in\cS$ is
arbitrary. There are three additional families of
algebraic spin eigenfunctions with zero total momentum, given by
\begin{equation*}
\begin{aligned}
\Psi_{1,2}^{(1)}&=\mu\sum_i\bigg\{\begin{matrix}\cos\\\sin\end{matrix}\bigg\}
\big(2(x_i-X)\big)\ket{s_i}\,,\\
\Psi_3^{(1)}&=\mu\bigg[\frac
a{2a+1}\sum_i\ket{s_i}+\sum_{i<j}\cos(x_i-x_j)\ket{s_j}\bigg]\,,
\end{aligned}
\end{equation*}
where the spin states $\ket{s_i}$ are defined by $\La\big(z_1\ket
s\big)=\sum_iz_i\ket{s_i}$, and $\ket s\in\cS$ is any non-symmetric state. It should
be noted that each of the spin states $\Psi^{(0)}_n$ and $\Psi^{(1)}_n$
has the same energy as the scalar eigenfunction $\psi_n$.

\emph{Case 2.} In this case the algebraic energies are given by
$E_n=E_0+nc_0=E_0+4n\omega$, where the quantum number
$n=0,1,\dots$ is the degree in $\bz$ of the corresponding
eigenfunction of the gauged Hamiltonian $\tH_2$. This follows
easily from the fact that both $T_2^*$ and $J_-$ lower the degree,
while $J_0$ preserves it. As for Calogero's original model, the
algebraic spectrum is equally spaced, but the spacing is twice the value
suggested by the harmonic term. Unlike the usual CS models,
the algebraic levels of this model have a well-defined thermodynamic limit, i.e.,
$E_n/N\to\omega(4a+2b+1)$ as $N\to\infty$. This property, which was
already noted in Ref.~\cite{JK99} for the scalar reductions of the potentials~\eqref{V00}
and~\eqref{V--}, is in fact shared by all the models~\eqref{Vs}. In the scalar case, for
each $n\geq 2$ there are two algebraic eigenfunctions with energy
$E_n$, namely
\begin{align*}
\psi^{(0)}_n={} &\mu\ms L_n^{\al-1}(\omega r^2),\qquad n=0,1,\dots\,,
\\
\psi^{(1)}_n={} &\mu\Big[\omega^2\Big(N(\al+1)\sum_i x_i^4-\al r^4\Big)\,L_{n-2}^{\al+3}(\omega r^2)
\hspace{5em}
\\
&\hspace{9em}{}+n(n-1)\al\,L_n^{\al-1}(\omega r^2)\Big],\qquad n=2,3,\dots\,,
\end{align*}
where $\al=N(2a+b+1/2)$ and $L^\la_\nu$ is a generalized Laguerre
polynomial of degree $\nu$. Note, in particular, that
$\psi^{(0)}_0=\mu$ has no nodes in the configuration space
$0<x_1<\cdots<x_N$, and is thus the ground state wavefunction. In
the spin case, for each energy level $E_n$ we have first of all
the factorized eigenfunctions of the form
$\Psi^{(k)}_n=\psi^{(k)}_n\La\ket s$, $k=0,1$.
In addition, for each $n\geq 1$ there is a family of
genuine spin eigenfunctions
\begin{equation*}
\Psi^{(2)}_n=\mu\Big[L^{\al+1}_{n-1}(\omega r^2)\La\big(x_1^2\ket s\big)-\frac\al{N\omega}\,
L^{\al}_{n-1}(\omega r^2)\La\ket s\Big],
\end{equation*}
where $\ket s\in\cS$ is non-symmetric.

\emph{Case 3.} This is probably the most interesting case, since
the algebraic eigenfunctions depend essentially on the three
symmetric variables $\tau_1$, $\tau_2$, $\tau_3$. As in the previous case,
the algebraic energies are given by $E_n=E_0+nc_0=E_0+2n\omega$, where
$n$ is again the degree in $\bz$ of the corresponding eigenfunctions of $\tH_3$.
We shall begin, as usual, with the scalar case, for which we have been able to
compute all the algebraic eigenfunctions in closed form.
For each energy level $E_n$ with $n\geq 3$ there are two
infinite families of algebraic eigenfunctions
$\psi^{(k)}_{lm}$, $k=0,1$, with $n=2l+m$.
The first family is given by
\begin{equation}\label{psi0lm}
\psi^{(0)}_{lm}=\mu\tau_1^mL^{-\be}_l(\omega r^2)
P^{(\al,\be)}_{[\frac m2]}(t)\,,\quad l,m=0,1,\dots,
\end{equation}
where $t=2Nr^2/\tau_1^2-1$, $\al=N(a+1/2)-3/2$,
$\be\equiv\be(m)=1-m-N(a+1/2)$, $P^{(\al,\be)}_\nu$ is a
Jacobi polynomial of degree $\nu$, and $[x]$ denotes the integer part of
$x$. The second family reads
\begin{equation}\label{psi1lm}
\begin{split}
\psi^{(1)}_{lm}={}&\mu\tau_1^{m-3}L^{-\be}_l(\omega r^2)\bigg[
P^{(\al+3,\be)}_{[\frac{m-3}2]}(t)\sum_i x_i^3+\frac{3\tau_1^3}{2N^2}\,\vp_{lm}(t)\bigg],
\quad l,m-3=0,1,\dots,
\end{split}
\end{equation}
where $\vp_{lm}$ is a polynomial of degree $[m/2]$ given by
\begin{multline}\label{vplmeven}
\vp_{lm}=\frac{m+2\al+2}{m-1}\,P^{(\al+2,\be-2)}_{\frac m2}(t)
-P^{(\al+3,\be-1)}_{\frac m2-1}(t)\\
-\frac{4\al+7}{m-1}\,P^{(\al+2,\be-1)}_{\frac m2-1}(t)
+\frac13\,P^{(\al+3,\be)}_{\frac m2-2}(t)\,,
\end{multline}
for even $m$, while for odd $m$ we have
\begin{multline}\label{vplmodd}
\vp_{lm}=2P^{(\al+2,\be-1)}_{\frac{m-1}2}(t)-P^{(\al+3,\be-1)}_{\frac{m-1}2}(t)
+\frac13\,P^{(\al+3,\be)}_{\frac{m-3}2}(t)\\
+\frac{m+2\al+2}{m(m-2)}\,P^{(\al+1,\be)}_{\frac{m-3}2}(t)
-\frac{m+2\al+2}{m-2}\,P^{(\al+2,\be)}_{\frac{m-3}2}(t)\,.
\end{multline}
Note that, as in the previous cases, $\psi_0^{(0)}=\mu$ is the
ground state.
The above results show that for all $N\geq 3$
the scalar reduction of the Hamiltonian $H_3$ possesses two different
families of eigenfunctions of the form
$\mu L_l^{-\be}(\omega r^2)p_\nu(\bx)$, with $p_\nu$ a homogeneous polynomial
of degree $\nu\geq 3$, cf.~Eqs.~\eqref{psi0lm} and \eqref{psi1lm}--\eqref{vplmodd}.
This was verified only up to $\nu=6$ and $N\geq\nu$ in Ref.~\cite{AJK01}.

In the spin case, the algebraic energies are of course the same as
in the scalar case. As in the previous cases, each scalar
algebraic eigenfunction gives rise to $\binom{2M+N}N$ factorized
spin eigenfunctions. The computation of the remaining algebraic
spin eigenfunctions, which is considerably more involved than in
Cases 1 and 2, is still in progress. Even without an explicit
knowledge of the eigenfunctions, for each algebraic level $E_n$
one can easily compute the number of independent states of the
form~\eqref{cT3} of degree $n$. If all
the eigenfunctions of this model (and of its scalar reduction) were
algebraic, the previous remark would imply that the
degeneracy of the levels can be explicitly found. The method of
Ref.~\cite{EFGR04} could then be used to compute the partition
function of the associated spin chain
\[
\ssH_3=\sum_i(\xi_i-\xi_{i+1})^{-2}\ms S_{i,i+1},
\]
where $(\xi_1,\dots,\xi_N)$
is an equilibrium of the scalar potential
\[
U_3=\frac12\,r^2+\sum_i\frac1{(x_i-x_{i-1})(x_i-x_{i+1})}+
\sum_i\frac1{(x_i-x_{i+1})^2}\,.
\]
This remark obviously applies to the rational
model~\eqref{V0-} as well. Note, however, that for this model one can also
construct states of the associated spin chain by applying the freezing
trick to the genuine spin eigenfunctions $\Psi^{(2)}_n$ presented above.
%
%
\begin{ack}
 This work was partially supported by the DGI under grant no.~BFM2002--02646.
 A.E. acknowledges the financial support of the Spanish Ministry of Education through
 an FPU scholarship.
\end{ack}

\end{document}